\input epsf

\documentstyle[epsfig]{aipproc2}

\def\ltsima{$\; \buildrel < \over \sim \;$}
\def\simlt{\lower.5ex\hbox{\ltsima}}
\def\gtsima{$\; \buildrel > \over \sim \;$}
\def\simgt{\lower.5ex\hbox{\gtsima}}

\begin{document}
\title{Are pre-big-bang models falsifiable by gravitational wave experiments?}

\author{Carlo Ungarelli\thanks{Present address: School of Computer 
Science and Mathematics, University of Portsmouth, Mercantile House, 
Hampshire Terrace, Portsmouth P01 2EG, UK} and Alberto Vecchio}
\address{Max Planck Institut f\"{u}r Gravitationsphysik,
Albert Einstein Institut\\
Am M\"{u}hlenberg 1, D-14476 Golm, Germany}

\begin{flushright}
PU-RCG/99/19
\end{flushright}
\maketitle

\begin{abstract}
One of the most interesting predictions of string-inspired
cosmological models is the presence of a stochastic background of
relic gravitational waves in the frequency band accessible to
Earth-based detectors.
Here we consider a ``minimal'' class of string cosmology models
and explore whether they are falsifiable
by gravitational wave observations. In particular, we show that,
the detectability of the signal
depends crucially on the actual values of the model parameters. This
feature will enable laser interferometers -- starting from the
second generation of detectors -- to place stringent constraints on
the theory for a fairly large range of the free parameters
of the model.
\end{abstract}

\section*{Motivations}

One of the direct phenomenological predictions of inflationary
cosmological models is the generation of a stochastic background of
gravitational waves (GW's). For ``slow-roll'' inflationary models this prediction 
is hardly testable: since the spectrum is almost flat over the huge 
frequency band $\sim 10^{-16}\,{\rm Hz} - 1\,{\rm GHz}$~\cite{GWInfl},
in the window $10\,{\rm Hz} - 1\,{\rm kHz}$, where ground-based GW experiments operate,
the maximum value of the energy spectrum $\Omega_{\rm gw}(f) \equiv \rho_c^{-1}
d\rho_{\rm gw}(f)/d\ln(f)$ compatible with the COBE 
measurements of the cosmic micro-wave background (CMB) temperature fluctuations 
at large scales -- $h_{0}^2\Omega_{\rm gw}\leq 10^{-14}$~\cite{KW92} --
is well below the sensitivity expected for the third generation of detectors,
$h_{0}^2\Omega_{\rm gw}\simgt 10^{-11}$.

``Pre-big-bang'' (PBB) models~\cite{PRBB} 
represent an alternative to the standard ``slow-roll'' inflationary scenario. In a
minimal PBB model, one assumes that the initial state is the perturbative 
vacuum of super-string theory, the flat 10-dimensions Minkowski space-time; the
Universe goes first through an inflationary phase where the curvature and the
string coupling  increase, eventually reaches a 
``stringy epoch'' where the curvature scale is of order one in units 
of the string length, and finally evolves into a
typical decelerated radiation/matter dominated era. The inflationary PBB phase
has a precise consequence on the structure of $\Omega_{\rm gw}(f)$, which affects 
the detectability of the stochastic background by laser interferometers, 
whose "science-runs" are expected to begin at the end of 2001: in the low frequency range, the spectrum is characterized by a steep
power law, $\Omega_{\rm gw}\propto f^3$~\cite{BGGV}; indeed the COBE bound is easily evaded,
and the spectrum can peak at frequencies of interest for GW experiments,
while satisfying the existing experimental bounds~\cite{BGV,BMU}. 

\begin{figure} 
\centerline{\epsfig{file=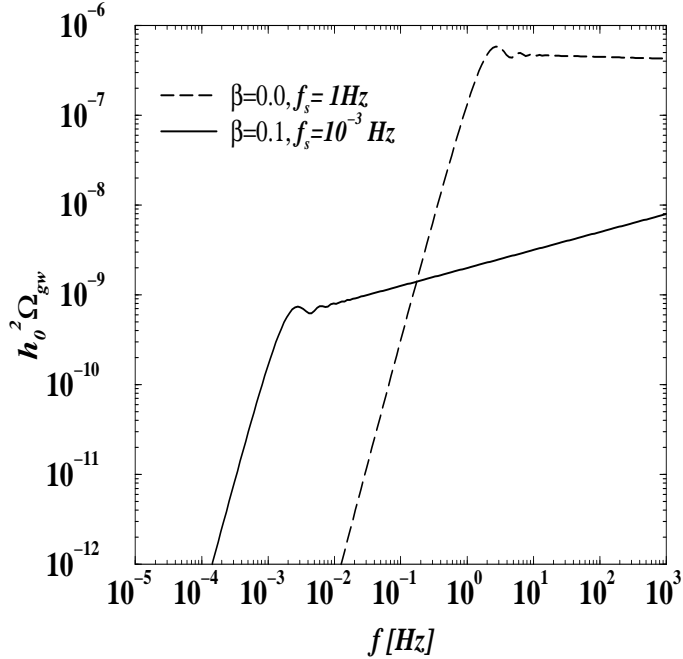,height=3.5in,width=3.5in}}
\vspace{10pt}
\caption{The energy spectrum $h_0^2\,\Omega_{\rm }(f)$ of relic gravitons
according to the prediction of minimal pre-big-bang models as a function of the 
frequency (in Hz)  for two different choices of the model parameters; here the
parameter $\beta$ is defined as $\beta=-\frac{\log(g_s/g_1)}{\log z_s}$.}
\label{fig1}
\end{figure}

For the rather general class of minimal PBB models, $\Omega_{\rm gw}(f)$
depends on two free parameters: (i) the red-shift $z_s \equiv f_1/f_s$ of the high-curvature 
phase, which fixes the ``knee'' frequency $f_s$ between the low 
and high frequency regime ($f_1\sim 10^{10}$ Hz is the cut-off
frequency of the spectrum, whose exact value is irrelevant for 
the issues discussed here); (ii) the value $g_s$ of the string
coupling at the onset of the high-curvature phase, which determines
the high-frequency slope of the spectrum. 
In Fig.~\ref{fig1} we show $\Omega_{\rm gw}(f)$ for two
choices of the free parameters; varying $g_s$ and $z_s$, the maximum
value of the spectrum compatible with the constraints due to pulsar
timing data and to the abundance of light elements at the nucleosynthesis
epoch is $h_0^2\Omega_{\rm gw} \sim 10^{-7}$~\cite{BMU}. %
$\Omega_{\rm gw}(f)$ strongly depends on $z_s$ and $g_s$: they affect both
the frequency behaviour and the peak value of the spectrum. Whether GW
experiments will be able or not to detect a signal predicted by PBB
models does depend on the actual value of the "true" parameters. Indeed, GW
experiments represent one of the very few avenues where the cosmological models can 
be verified, and could open a new era for studies of the very-early Universe 
and the structure of fundamental fields at high energies~\cite{Creighton,Maggiore}. 
The emphasis of this
contribution is not on finding the best theoretical scenario which would guarantee a detection.
The goal of our analysis is to address to which extent
GW observations can test PBB models, and more in detail to identify the 
the region of the free-parameter space that experiments can probe. A similar analysis,
in the context of string cosmology models, was carried out in~\cite{AB}.

\section*{Results}

In order to address whether a background characterized by a given energy spectrum
$\Omega_{\rm gw}(f; z_s,g_s)$ is indeed detectable or not, one needs to evaluate 
the signal-to-noise ratio
(SNR) that can be obtained by cross-correlating the output of two
detectors as a function of the model parameters, in our case $z_s$ and
$g_s$ (see also~\cite{AB}). The data analysis issues related to the detection of a stochastic
background have been thoroughly discussed in~\cite{ALRO}, 
where the reader can find full details. Here we recall only the
expression of the signal-to-noise ratio:
\begin{equation}
{\rm SNR} \approx \frac{3H_0^2}{10\pi^2}\, T_{\rm obs}^{1/2}
\left[\int_{-\infty}^\infty df\
\frac{\gamma^2 (|f|) \Omega_{\rm gw}^2(|f|; z_s,g_s)}{f^6 S_1(|f|) S_2(|f|)}
\right]^{1/2}\,.
\label{SNR_opt}
\end{equation}
In Eq.~(\ref{SNR_opt}) $T_{\rm obs}$ is the observation time -- which is usually assumed to
be a few months long -- $H_0=100\,h_0$ km sec$^{-1}$ Mpc$^{-1}$ is the Hubble constant,
$S_{i=1,2}(f)$ is the  noise spectral
density of the $i-$th detector, and $\gamma (f)$ is the overlap
reduction function. The spectrum $\Omega_{\rm gw}(f; g_s,f_s)$ for minimal
PBB models is given in~\cite{BMU}.

\begin{figure} 
\begin{center}
\centerline{\epsfig{file=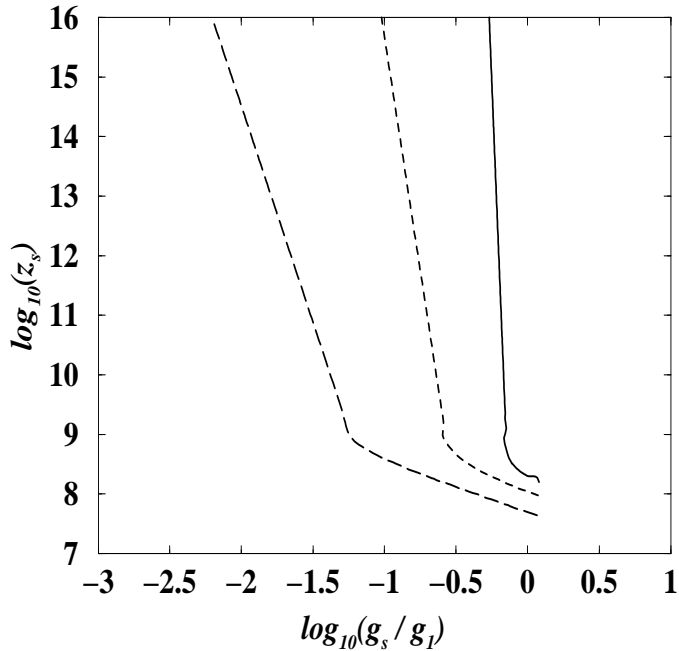,height=3.5in,width=3.5in}}
\caption{The signal-to-noise ratio (SNR) produced by the GW stochastic background 
predicted by minimal pre-big-bang models. 
The figure shows the contour plots of the SNR in the plane
of the free-parameters of the model, $f_s$ and $g_s$, obtained by cross-correlating the
two LIGO interferometers. The curves corresponds to: SNR = 0.1 for the initial
LIGO configuration (solid line); SNR = 5 for the enhanced LIGO configuration (short-dashed
line); SNR = 10 for the advanced LIGO configuration (long-dashed line). We remind
that the physical range over which the parameters are defined is:
$0 < g_s/g_1 < 1$ and $1 < z_s < 10^{16}$.
}
\label{fig2}
\end{center}
\end{figure}

Fig.~\ref{fig2} summarizes the results: it shows the
contour plots of SNR for $T_{\rm obs}=10^7$sec which can be achieved
by cross-correlating the two 4-km LIGO interferometers located in Hanford
and Livingston. The contours are drawn in the relevant 2-dimensional parameter
space: in order to highlight the physical content, our parameter
choice corresponds to $\log_{10}(z_s)$ and $\log_{10}(g_s/g_1)$; here
$g_1^2/4\pi=\alpha_{GUT}$, where $\alpha_{GUT}$ is the gauge coupling at the
unification scale ($\alpha_{GUT}\sim 1/20$). The two free-parameters
of the models are defined over the following range: $0 < g_s/g_1 < 1$ and $1 < z_s < 10^{16}$.
In our analysis we have considered the estimated noise
spectral density for the initial, enhanced and advanced LIGO configuration
(the so-called LIGO I, II, and III, respectively).

For initial LIGO, the minimum value of a detectable stochastic background is
$h_0^2\Omega_{\rm gw}^{\rm min}\sim 5\times 10^{-6}$, cfr. also~\cite{ALRO}, and therefore
the experiments can not provide any direct hint about PBB models: the maximum value of SNR,
for fined-tuned parameters of the model, is just below one. However, 
the enhanced LIGO configuration  will allow us to explore a relatively large parameter region: for SNR $> 5$, one can
probe models where $g_s$ differs by one order of magnitude with respect to its present value $g_1$ and about eight orders of magnitude in $f_s$; 
the SNR reaches a maximum $\sim 100$ when $g_s\simeq g_1$. 
The advanced LIGO configuration -- which however requires major technological developments --
will enlarge the $g_s$-range that one can test by
more than one order of magnitude; for selected parameter values the experiment
could reach SNR $\approx 1000$. %
Notice that the experiments are very sensitive to $g_s$, and achieve the maximum
SNR for $g_s \sim g_1$. 

One might wonder whether other interferometers and/or resonant detectors could
provide relevant information for the problem at hand.
Unfortunately the location, orientation and sensitivity of VIRGO, GEO600 and TAMA are
such that any search involving one of these three instruments 
will not allow to detect a stochastic background of PBB gravitons; just as 
reference, a VIRGO-GEO600 cross-correlation experiment will reach
$h_0^2\Omega_{\rm gw}^{\rm min}\sim 8\times 10^{-6}$ for an integration
time of 4-months. Neither resonant antenna are suitable: two "bars" operating at the quantum
limit with exactly the same resonance frequency would reach a maximum sensitivity
$h_0^2\Omega_{\rm gw}^{\rm min}\sim 5\times 10^{-6}$ (see also~\cite{Vitale}),
which is a factor $\sim 10$ worse than the one required for the most
optimistic theoretical prediction; experiments involving hollow spheres~\cite{Coccia},
which are currently under study, could reach $h_0^2\Omega_{\rm gw}^{\rm min}\sim 10^{-7}$
for detectors operating at the quantum limit and co-located;
this would produce a SNR $\sim 1$ for a PBB model whose parameters are such that
the peak of $\Omega_{\rm gw}(f)$ is right in the middle of the sphere frequency band.

\section*{Conclusions}

Minimal PBB models predict a GW stochastic background which could be detectable 
by cross-correlating the two LIGO instruments -- or, more in general, two 
interferometers at a distance $< 3000$ km and quasi-optimally oriented -- 
with a sensitivity which is intermediate between the first and second
stage. In the time frame 2004-2005, when the detectors are expected to operate in the
so-called enhanced configuration, we will be able to place experimental constraints
on PBB models for a fairly large portion of the free-parameter space. A more 
detailed analysis regarding the issues discussed here is currently in preparation.


\begin{references}
\bibitem{GWInfl}
B.~Allen, {\it Phys. Rev D} {\bf 37}, 2078 (1988). 
A.~Liddle and M.~Turner, {\it Phys. Rev. D} {\bf 50}, 758 (1994); 
M.~Turner, {\it Phys. Rev. D} {\bf 55}, 435 (1997).
\bibitem{KW92}
L.~Krauss and M.~White, {\it Phys. Rev. Lett.} {\bf 69}, 869 (1992).
\bibitem{PRBB}
G.~Veneziano, {\it Phys. Lett. B}{\bf 265}, 287 (1991);
M.~Gasperini and G~Veneziano, {\it Astropart. Phys.} {\bf 1}, 317 (1993), 
{\it Mod. Phys. Lett. A}{\bf 8}, 3701 (1993). 
\bibitem{BGGV}
R. Brustein, M. Gasperini, M. Giovannini and G. Veneziano,
{\it Phys. Lett. B} {\bf 361}, 45 (1995).
\bibitem{BGV}
R. Brustein, M. Gasperini and G. Veneziano, {\it Phys. Rev. D} {\bf 55}, 3882 
(1997).
\bibitem{BMU}
A. Buonanno, M. Maggiore, C. Ungarelli, {\it Phys. Rev. D} {\bf 55}, 
3330 (1997).
\bibitem{Creighton}
T.~Creighton, preprint gr-qc/9907045.
\bibitem{Maggiore}
M.~Maggiore, to appear in Physics Reports, preprint gr-qc/9909001.
\bibitem{AB}
B. Allen and R. Brustein, {\it Phys. Rev. D} {\bf 55}, 3260 (1997).
\bibitem{ALRO}
B. Allen and J.D. Romano, {\it Phys. Rev. D} {\bf 59}, 102001 (1999).
\bibitem{Vitale}
S. Vitale, M. Cerdonio, E. Coccia and A. Ortolan {\it Phys. Rev. D} 
{\bf 55}, 1741 (1997). 
\bibitem{Coccia}
E. Coccia, V. Fafone, G. Frossati, J. A. Lobo and J. A. Ortega 
{\it Phys.Rev. D} {\bf 57}, 2051 (1998). 
\end{references}
\end{document}